\documentclass[preprint,prx,amsmath,amssymb,longbibliography,superscriptaddress]{revtex4-1}

\usepackage{graphicx}
\usepackage{dcolumn}
\usepackage{bm}
\usepackage{amsmath,latexsym,tabularx}
\usepackage{multirow}

\usepackage{hyperref}
\hypersetup{
  colorlinks=true,
  urlcolor=blue,
  linkcolor=blue,
  citecolor=blue
}

\usepackage{multirow,booktabs}
\usepackage{comment}

\newcommand{\Sr}{$^{88}$Sr$^+$}

\newcommand{\figRef}[1]{\mbox{Fig. \ref{#1}}}

\newcommand{\affilLL}[0]{Lincoln Laboratory, Massachusetts Institute of Technology, Lexington, Massachusetts 02421, USA}
\newcommand{\affilMIT}{Massachusetts Institute of Technology, Cambridge, Massachusetts 02139, USA}

\begin{document}

\title{Integrated multi-wavelength control of an ion qubit}

\author{R.~J.~Niffenegger}
\email[]{robert.niffenegger@ll.mit.edu}
\affiliation{\affilLL}

\author{J.~Stuart}
\affiliation{\affilLL}
\affiliation{\affilMIT}

\author{C.~Sorace-Agaskar}
\affiliation{\affilLL}

\author{D.~Kharas}
\affiliation{\affilLL}

\author{S.~Bramhavar}
\affiliation{\affilLL}

\author{C.~D.~Bruzewicz}
\affiliation{\affilLL}

\author{W.~Loh}
\affiliation{\affilLL}

\author{R.~T.~Maxson}
\affiliation{\affilLL}

\author{R.~McConnell}
\affiliation{\affilLL}

\author{D.~Reens}
\affiliation{\affilLL}

\author{G.~N.~West}
\affiliation{\affilMIT}

\author{J.~M.~Sage}
\email[]{jsage@ll.mit.edu}
\affiliation{\affilLL}
\affiliation{\affilMIT}

\author{J.~Chiaverini}
\email[]{john.chiaverini@ll.mit.edu}
\affiliation{\affilLL}
\affiliation{\affilMIT}


\date{\today}

\begin{abstract}
Monolithic integration of control technologies for atomic systems is a promising route to the development of quantum computers and portable quantum sensors~\cite{mehta2016integrated,slichter2017uv,todaro2020state,PhysRevApplied.11.024010}. Trapped atomic ions form the basis of high-fidelity quantum information processors~\cite{PhysRevLett.74.4091,bruzewicz2019trapped} and high-accuracy optical clocks~\cite{brewer2019clock}.  However, current implementations rely on free-space optics for ion control, which limits their portability and scalability. Here we demonstrate a surface-electrode ion-trap chip~\cite{NIST:SET:QIC:05,PhysRevLett.96.253003} using integrated waveguides and grating couplers, which delivers all the wavelengths of light required for ionization, cooling, coherent operations, and quantum-state preparation and detection of Sr$^{+}$ qubits. Laser light from violet to infrared is coupled onto the chip via an optical-fiber array, creating an inherently stable optical path, which we use to demonstrate qubit coherence that is resilient to platform vibrations. This demonstration of CMOS-compatible integrated-photonic surface-trap fabrication, robust packaging, and enhanced qubit coherence is a key advance in the development of portable trapped-ion quantum sensors and clocks, providing a way toward the complete, individual control of larger numbers of ions in quantum information processing systems.
\end{abstract}

\maketitle

Trapped ions are a promising technology for quantum information processing, owing to their fundamental reproducibility and ease of control and readout, and because their long coherence times and strong ion-ion interactions enable high-fidelity two-qubit gates~\cite{leibfried2003quantum,monroe2013scaling, gaebler2016highfidelity,bruzewicz2019trapped,PhysRevLett.117.060504}. Quantum processors based on trapped ions have been used to demonstrate relatively complex quantum algorithms in architectures with high connectivity~\cite{nam2019ground, wright2019benchmarking}.  Trapped ions can also be used as precise quantum sensors and as the basis for high-accuracy optical clocks~\cite{PhysRevX.7.031050,delahaye_transportable_ion_clocks_2018, brewer2019clock}.
There are nevertheless many challenges to increasing the number of trapped-ion qubits in a quantum processor while maintaining high-fidelity operations, and to developing portable trapped-ion-based quantum sensors.  
Chief among these are the numerous free-space optical elements used to tightly focus and direct multiple laser beams of varied wavelength to the location of each ion. Optical beam paths defined by these elements are susceptible to vibrations and drift, causing beam pointing instability, which can limit the fidelity of quantum logic operations in quantum computers and the sensitivity of quantum sensors deployed outside the laboratory.  This limitation can be direct, through intensity fluctuations~\cite{Thom2018}, or indirect, through the requirement to use larger beams and hence smaller detunings from resonance.  The indirect limitation may result in an unfavorable trade-off between spontaneous-emission error and power requirements~\cite{gaebler2016highfidelity,PhysRevLett.117.060504}.

Photonic waveguides and grating couplers~\cite{Dalgoutte:75} integrated into microfabricated surface-electrode ion traps~\cite{NIST:SET:QIC:05} offer a way to overcome the limitations of free-space optics for light delivery to trapped ions, greatly reducing the complexity and experimental overhead of trapped-ion systems. A chip-integrated single-mode waveguide has allowed coherent operations on a trapped ion~\cite{mehta2016integrated}, and a multi-material photonics platform has been developed to enable low-loss light delivery over the wavelength range relevant to commonly used ion species (such as Ca$^{+}$\!, Sr$^{+}$\!, Ba$^{+}$\!, and Yb$^{+}$)~\cite{sorace2019versatile}. However, the monolithic integration of optical components for the delivery of all light required for basic ion-qubit operations has remained an outstanding challenge.

Here we demonstrate the operation of a surface-electrode ion-trap chip in which integrated photonic components deliver all wavelengths (violet to infrared) necessary to control \Sr\ qubits.
Using these integrated components, we demonstrate operations including photoionization of neutral Sr, Doppler and resolved-sideband cooling of the \Sr\ ion, electronic-state repumping, coherent qubit operations, and qubit state preparation and detection.
In contrast to previous demonstrations, in which the input beams were coupled to the chip from free space~\cite{mehta2016integrated}, in our demonstration all laser wavelengths are coupled onto the chip via optical fibers mounted in a glass fiber-array block that is bonded to the chip.  Our demonstration thus markedly reduces the number of free-space optics for delivery of control light to the ion.
We observe substantially reduced sensitivity of ion qubits to external vibrations when compared to free-space-optics-based light delivery. 
The operation of this multi-wavelength photonics ion trap is a critical step toward robust and portable trapped-ion-based clocks and quantum sensors It also demonstrates the promise of this technology for ion-array-based quantum information processing with many qubits.


\begin{figure*}[ht!]
\centering
\includegraphics[width=0.9\textwidth]{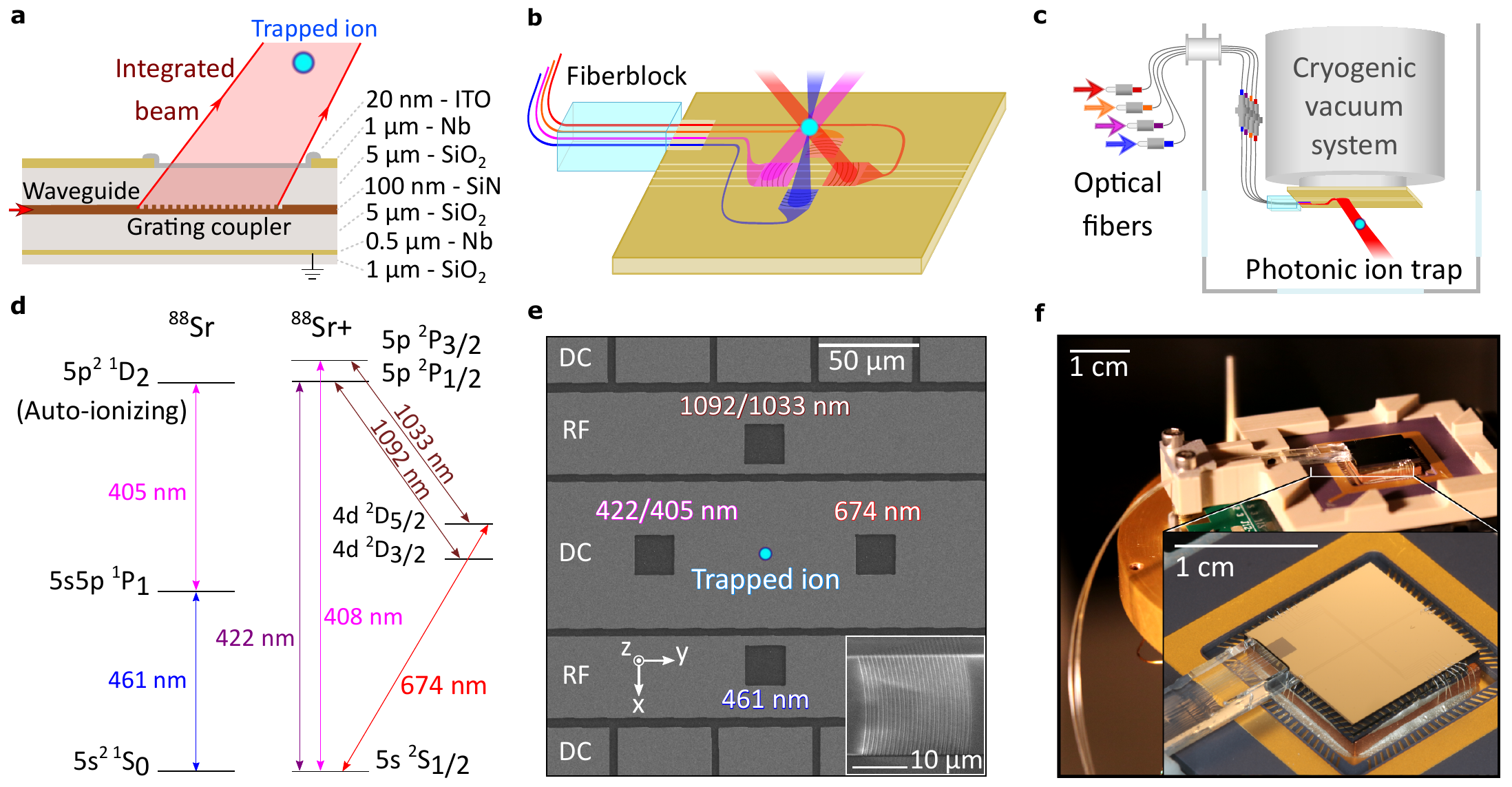}
\caption{Ion-trap-integrated photonic elements and experimental setup. 
(a) Photonic ion-trap cross section (not to scale), showing the integrated waveguides beneath the surface-trap electrodes and light diffracting out of the chip by means of a grating coupler.
(b) The fibers are arranged in a fiber-array block, which is aligned to photonic waveguides within the chip. The waveguides route the light to the center of the chip where vertical grating couplers diffract the light toward the ion trapped above the surface of the electrodes (the paths depicted are only notional).
(c) Light is coupled to the integrated photonic trap chip via optical fibers, which enter the cryogenic vacuum chamber through a fiber feedthrough.
(d) Energy-level diagram for Sr and Sr$^{+}$, depicting the wavelengths necessary for ion loading and control. 
(e) Scanning-electron micrograph (SEM) of the center of the trap showing square windows in the electrodes for the underlying grating couplers.
Inset, SEM of a grating coupler during fabrication showing the curvature of the grating lines, leading to transverse beam focusing.  RF, radiofrequency.
(f) Photonic ion-trap chip packaged and mounted with strain relief.
Inset, close-up of the 1~cm~$\times\ $1~cm chip, without full epoxy for clarity.
}
\label{fig:setup}
\end{figure*}

A schematic depiction of our ion-trap chip with integrated photonics is shown in \figRef{fig:setup}a, b.  Laser light is coupled into four fibers and delivered into the cryogenic ultra-high-vacuum (UHV) system by means of a custom fiber feed-through (\figRef{fig:setup}c).  Cryogenic operation is chosen primarily for the convenience of quickly obtaining UHV via cryopumping; it does not affect the basic functionality of the trap or the integrated photonic components.
Inside the vacuum chamber, the fibers terminate in a fiber array that is aligned and bonded to the chip's facet.  Light is coupled from these fibers onto the chip via inverse-taper waveguide couplers, sections of waveguide that are narrowed such that the spatial mode more closely matches the optical fiber mode~\cite{Almeida:03}.  
Each integrated-photonic waveguide then routes light under the metal trap electrodes to a grating coupler that directs the light vertically, through apertures placed in the metal electrodes, toward the ion-trapping zone above the surface of the chip.

The optical waveguides consist of a laterally defined silicon nitride (SiN) guiding layer surrounded by silicon dioxide (SiO$_2$) cladding; each waveguide is patterned to have a width that ensures single-mode operation at its design wavelength (ranging from a width of $250$~nm for a design wavelength of 405~nm to a width of $1.1~\mu$m for a design wavelength of 1092~nm). These polarization-maintaining, single-mode waveguides allow flexible routing of light to arbitrary locations below the chip surface. 
Near the end of each waveguide, the width is tapered up to $18~\mu$m in order to expand the spatial mode of the light before it reaches the grating coupler. 

The diffractive grating couplers are created by etching a periodic pattern into the widened waveguide along the direction of propagation (\figRef{fig:setup}a). 
This creates a periodic variation of the effective refractive index, causing light to diffract out of the plane of the chip at an angle dependent on the wavelength, grating period, and effective index.  For the designs in this work, the grating efficiencies are approximately $10\%$ (see Methods).  To increase the light intensity at the ion location, the grating teeth are curved, focusing the beam to a few-micron waist in the direction parallel to the trap surface and perpendicular to the grating emission (\figRef{fig:setup}e, inset).

The linear-ion-trap electrode geometry is similar to that used previously~\cite{sage2012loading}; fabrication is detailed in Methods. A radiofrequency (RF) drive applied to two of the trap electrodes confines the ion radially in the $x$ and $z$ directions, at a height of $z$=$55~\,\mu$m above the chip surface, while DC voltages applied to the other electrodes provide axial confinement along the $y$ direction and allow one-dimensional shuttling of the ion parallel to the trap surface. (See \figRef{fig:setup}e for an image of the trap electrodes and definition of the $x$, $y$, and $z$ axes.) Four $20~\mu$m~$\times\ 20~\mu$m apertures
are opened in the trap metal above the grating couplers.  A thin film of the transparent, conductive material indium tin oxide (ITO) is deposited over these apertures to reduce the exposed area of dielectric that could potentially become charged
~\cite{wang2011laser} and adversely affect the electric field at the ion location.  
After fabrication,
the edge of the chip, to which the inputs of the waveguides are routed, is polished in order to minimize light scattering at the interface between the waveguides and the optical fibers. The fiber array is then aligned and attached to the chip (see Methods).  

\begin{figure*}
\centering
\includegraphics[width=0.97\textwidth]{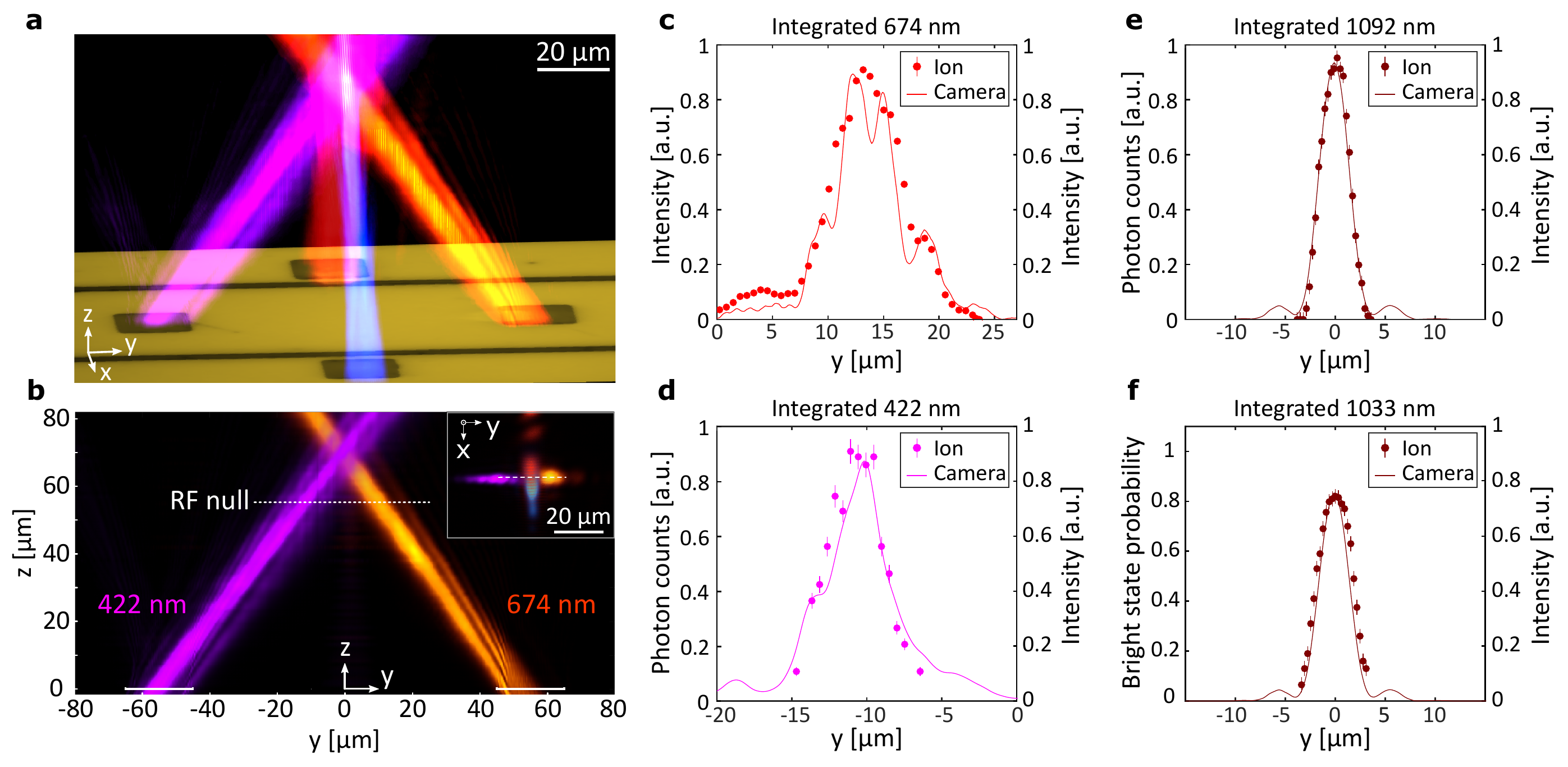}
\caption{Integrated photonic beam profiles measured via microscope and subsequently verified via ion interactions \textit{in situ}.
(a) False-color three-dimensional reconstruction of integrated beam profiles using images taken via a high-numerical-aperture microscope as its focal plane is scanned vertically above the chip.
(b) False-color 674~nm and 422~nm beam profiles in the $y$-$z$ plane (inset shows $x$-$y$ plane with all beams, $z=55$~$\mu$m).  (c) 674~nm integrated profile measured via Rabi oscillations at different positions and line cut of the beam intensity from microscope measurements (line). Profiles for 422~nm (d) and 1092~nm (e) measured via ion fluorescence.
(f) 1033~nm integrated beam profile measured via qubit quenching. Error bars indicate the standard error of the mean.  a.u., arbitrary units.
}
\label{fig:profile}
\end{figure*}

Six wavelengths are needed for loading and control of \Sr (\figRef{fig:setup}d). 
Ion loading is achieved by photoionization of neutral Sr from a remote, precooled source~\cite{sage2012loading,bruzewicz2016scalable} and requires 461-nm and 405-nm light.
Doppler cooling and detection on the \Sr\ $S_{1/2} \rightarrow P_{1/2}$ cycling transition require 422-nm light, as well as 1092-nm light for repumping from the $D_{3/2}$ state.  Coherent qubit operations, qubit state preparation via optical pumping into a single Zeeman sublevel of the $S_{1/2}$ state, and resolved-sideband cooling are performed by driving $S_{1/2} \rightarrow D_{5/2}$ electric quadrupole transitions via 674-nm light; light at 1033~nm is also used during optical pumping and resolved-sideband cooling.  The chip contains four separate photonics pathways (waveguides plus grating couplers) for ion control: one for 405-nm/422-nm light, one for 461-nm light, one for 674-nm light, and one for 1033-nm/1092-nm light.  Optical loss was determined for each pathway through a series of auxiliary measurements (see Methods).  For the cases where two different wavelengths propagate in the same waveguide (405~nm/422~nm and 1033~nm/1092~nm), the different wavelengths result in slightly different angles of emission from the grating couplers. Because the beams are focused only in the horizontal transverse direction, parallel to the chip surface (e.g. the $y$ direction for light propagating in the $x$ direction before scattering from the grating; cf. Figs.~\ref{fig:setup} and~\ref{fig:profile}), they are sufficiently large that both address the ion.  Light emitted by the grating couplers was in all cases linearly polarized parallel to the surface of the chip (see Methods).  A quantization axis was defined via the application of a magnetic field of $4.3$~G normal to the trap-chip surface.


We initially characterized each grating coupler by directly profiling its emitted beam with a high numerical aperture (NA) microscope objective and projecting the beam onto a CMOS detector~\cite{mehta2016integrated,mehta2017precise} (see Methods). We used these images to generate 3D profiles of the beams and to determine the precise beam positions relative to the trap electrodes (\figRef{fig:profile}a). These profiles show that the different beams intersect each other at a height of 65~$\mu$m above the chip surface, 10~$\mu$m above the chip's RF null, the line along which the ion is nominally trapped (\figRef{fig:profile}b). 
This offset is consistent with a discrepancy between the index of refraction value used during grating design and that measured for the low-loss SiN used in fabrication (see Methods for further explanation). 

We next characterized the profiles, as well as the positions of the beams relative to the trap electrodes, \textit{in situ} using the ion, by measuring the strength of the laser-ion interaction as a function of ion position. The location of the ion was varied by changing the voltages applied to the DC electrodes (cf. Fig.~\ref{fig:setup}e), shuttling the ion along the direction of axial symmetry of the trap ($y$), in steps, through each beam (\figRef{fig:profile}c--f). For the 674-nm beam (\figRef{fig:profile}c), the frequency of Rabi oscillations on the $|S_{1/2},m_{J}=-1/2\rangle\rightarrow|D_{5/2},m_{J}=-5/2\rangle$ qubit transition was used to determine the beam intensity as a function of ion position. To characterize the 422-nm and 1092-nm beams, ion fluorescence as a function of ion position was used (\figRef{fig:profile}d and  \figRef{fig:profile}e, respectively).  
The integrated 1033-nm beam, emitted from the same coupler as the 1092-nm beam, was profiled by
applying a short 1033-nm pulse to partially quench the dark state population to the bright state (\figRef{fig:profile}f) (see Methods). These ion interaction profiles agree very well with the beam profiles measured using the microscope objective.

In addition to spatially profiling the integrated beams, we used them to implement several key ion-control operations.  Because we cannot move the ion vertically to the point where all integrated beams intersect, we instead demonstrate operations using up to three integrated beams at a time, shuttling the ion horizontally via the on-chip electrodes to the location of the relevant beam(s). Ion loading via photoionization was achieved within a few seconds using the integrated 461-nm beam and a free-space 405-nm beam, and in approximately 1~minute via the integrated 405-nm beam and free-space 461-nm beam. The longer loading time of the latter results from the broad 405-nm transition being unsaturated and from a lower 405-nm beam intensity relative to that available when using the free space beam.  We demonstrated quenching of the $D_{5/2}$ state with the integrated 1033-nm beam in less than 10~$\mu$s, and repumping from the $D_{3/2}$ state via the integrated 1092-nm beam for Doppler cooling and detection.  We used the integrated 674-nm beam path to perform spectroscopy on the qubit transition, as well as quantum state preparation (including optical pumping and resolved-sideband cooling to $\bar{n}<1$ motional quanta in the axial mode), Rabi oscillations, and Ramsey interferometry. With 10~mW of optical power coupled into the fiber attached to the chip, we were able to perform $\pi$~pulses on the optical qubit in 6.5~$\mu$s, consistent with measured losses (see Methods).

We implemented ion-control operations using combinations of integrated beams simultaneously by positioning the ion near the intersection of two or more beams, although this effectively reduced the available laser intensity for various operations. Doppler cooling and state detection with the integrated 422-nm and 1092-nm beams were performed, enabling trap lifetimes exceeding multiple hours and measurement of the ion state without the use of free-space beams.  For ion detection, 422-nm photons emitted from the ion are collected using a high-NA lens and counted via a photomultiplier tube.

\begin{figure}
\centering
\includegraphics[width=0.48\textwidth]{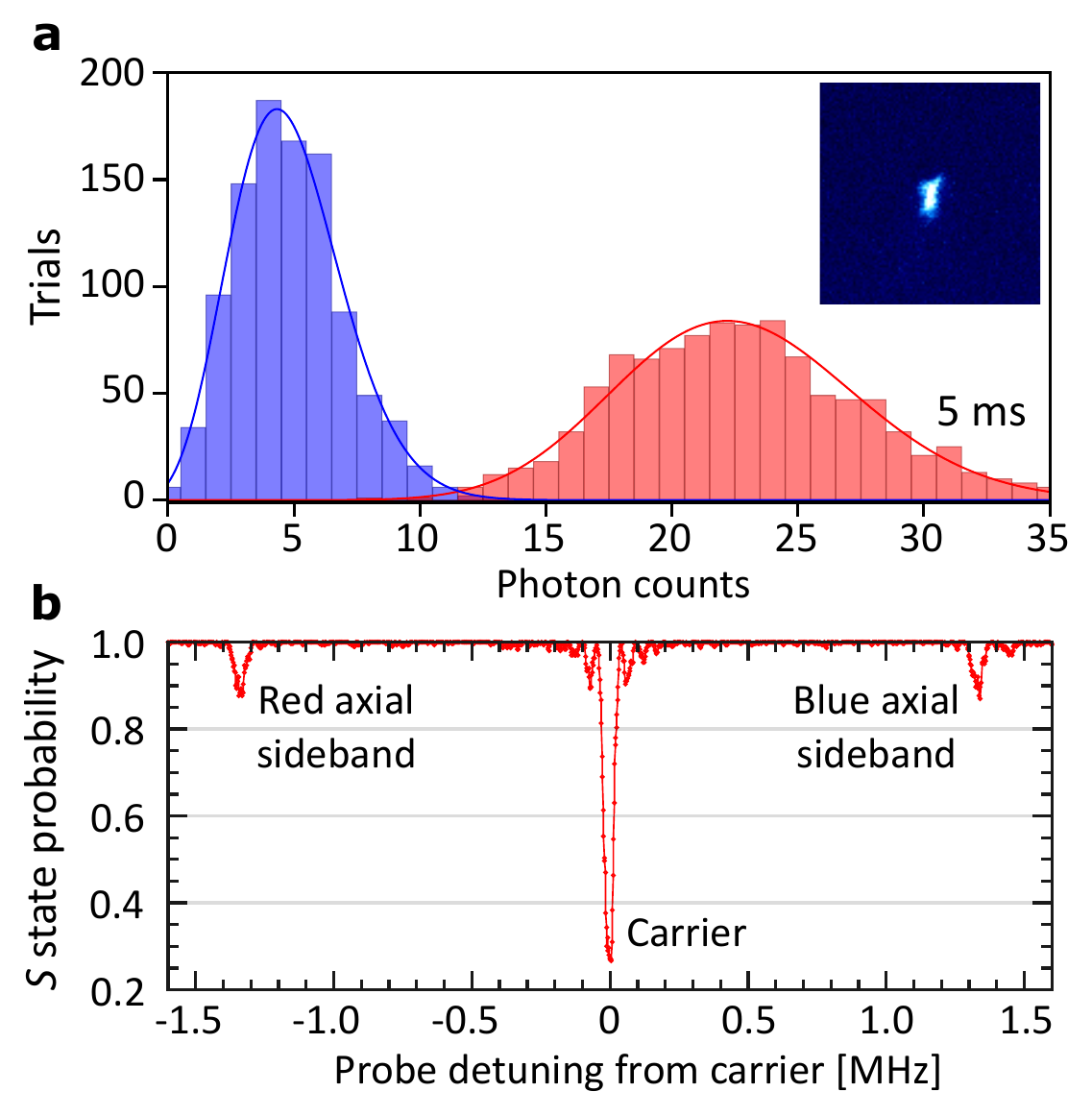}
\caption{Ion state detection and spectroscopy with integrated light delivery. (a) Histogram of bright state photon counts with the ion simultaneously illuminated by integrated 422-nm and integrated 1092-nm beams (red), and dark state/background photon counts without 1092-nm repumping light (blue), after a detection time of 5~ms (99.0\% expected mean qubit detection fidelity; see text).  Lines are fits to Poisson distributions. Inset, charge-coupled device (CCD) image of the ion illuminated by integrated beams.
(b) Ion spectroscopy and optical pumping, with 674-nm, 1033-nm, and 1092-nm light all delivered via integrated pathways, showing the $|S_{1/2},m_{J}=-1/2\rangle\rightarrow|D_{5/2},m_{J}=-5/2\rangle$ qubit carrier transition and the motional sidebands offset in frequency by the axial trap frequency (1.3~MHz).  
}
\label{fig:detection}
\end{figure}

\figRef{fig:detection}a shows a histogram of bright state ($S_{1/2}$) counts obtained with the ion simultaneously illuminated by integrated 422-nm and integrated 1092-nm beams for 5~ms, and counts from when the 1092-nm beam is blocked (causing the ion to be shelved into the dark $D_{3/2}$ state), equivalent to background during the same interval.  We did not prepare the ion in the $D_{5/2}$ state as part of this experiment, as there would be additional error due to state preparation.  However, if we assume the only additional optical-qubit detection error arises from spontaneous decay from the $D_{5/2}$ state (with lifetime of ${\sim}390$~ms~\cite{safronova2017forbidden}) during the measurement interval, then the expected mean qubit detection fidelity, evaluated as $1-(\epsilon_{d}+\epsilon_{b})/2$, is $99.0\%$ in 5~ms, where $\epsilon_{d}$ $(\epsilon_{b})$ is the probability to measure a dark (bright) ion as bright (dark).  Here, the error terms are calculated from the overlap of Poissonian fits to the ion-fluorescence and background-count histograms, including in the latter the expected counts for decay at various times during the detection interval, weighted by the probability to decay at any particular time (see Methods).  We also used the integrated 674-nm, 1033-nm, and 1092-nm beams simultaneously to perform qubit state preparation via optical pumping and spectroscopy of the $S_{1/2} \rightarrow D_{5/2}$ carrier transition and its first-order motional sidebands (\figRef{fig:detection}b).  We did not observe any noticeable effects of charging due to photo-liberated electrons from any of the integrated beams; variation in compensation voltages used to cancel stray fields (resulting in part-per-thousand trap-frequency variations over the course of a day) was comparable to that in traps without integrated photonics.

\begin{figure}[tbp]
\centering
\includegraphics[width=0.48\textwidth]{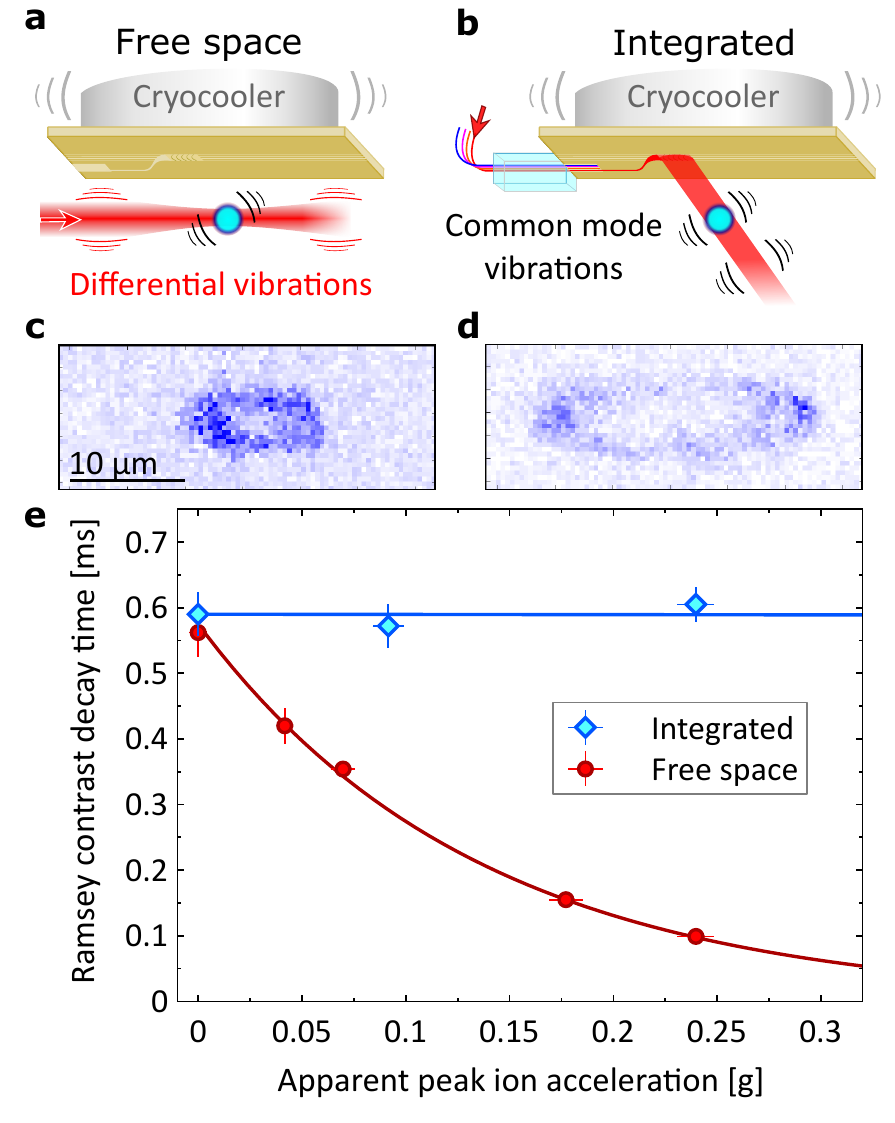}
\caption{ 
Vibration insensitivity when delivering qubit-control light via monolithically integrated optics and direct fiber-to-chip coupling.  
(a) Vibrations introduced from the cryocooler attached to the vacuum chamber cause the ion to vibrate relative to the free-space optical path. 
(b) Integrated photonic beams emitted from the chip will vibrate in common with the ion. 
(c,d) Long-exposure images of increasing vibrational coupling, collected using a high-numerical-aperture lens and an electron-multiplying CCD camera. Increasing the mechanical coupling increases the amplitude of ion oscillation, and hence ion acceleration for fixed oscillation period (we observe an oscillation period of approximately 15~ms). 
(e) The Ramsey-contrast decay time ($1/e$) measured via the free-space external beam path (red circles) decreases rapidly as the acceleration experienced by the vibrating ion increases. However, the coherence time measured when using delivery via the integrated beam path (blue diamonds) is unchanged.  Lines are exponential fits to the data in both cases and error bars indicate the standard error of the mean.
}
\label{fig:vibration}
\end{figure}

The inherent stability of optical paths integrated into the trap chip can provide vibration tolerance to trapped-ion qubits, clocks, and sensors.  Effective qubit decoherence due to optical phase variation and amplitude modulation arising from vibration of the ion with respect to the laser beam
(\figRef{fig:vibration}a) should be significantly reduced, because the vibrations of the ion and the light-delivery optics are common-mode, owing to monolithic integration with the trap (\figRef{fig:vibration}b).
To test this hypothesis, vibrations of varying amplitude were intentionally introduced, and we measured the effect on trapped-ion qubit coherence. 

We use a cryogenic vacuum system~\cite{sage2012loading,bruzewicz2016scalable} in which the cryocooler head is normally mechanically isolated from the trap mount.
By clamping the vibrating cryocooler head to the trap mount via the upper portion of the vacuum chamber, we can incrementally increase the vibrational coupling, causing the chip and ion to oscillate in space with a significant amplitude, as shown in~\figRef{fig:vibration}c-d.  
Here images of the ion fluorescence are observed as a function of time to determine the approximate amplitude of the induced vibration.

As a measure of qubit coherence, we used the decay of contrast of Ramsey interference fringes (see Methods). With the cryocooler head mechanically isolated from the trap, a $1/e$ contrast decay time of approximately 600~$\mu$s is measured using either the free-space 674-nm beam path or the grating-coupler-delivered 674-nm beam (see \figRef{fig:vibration}e, points at zero acceleration).  The contrast decay in these cases is limited by a combination of magnetic-field noise and uncompensated acoustic or thermal noise in the ${\sim}30$~m of fiber  used to transmit the light from the laser to the vacuum chamber.

Increasing the vibration coupling between the cryocooler and the trap mount as described above has a strong effect on the measured Ramsey-contrast decay when using the free-space qubit-control beam.  As shown in \figRef{fig:vibration}e, the Ramsey decay time as a function of the ion acceleration (as extracted from the ion motion; see Methods) decreases rapidly with increasing ion motion, owing to the vibration-induced randomization of the laser optical phase between the Ramsey pulses.  By contrast, the Ramsey decay measured using the grating-coupler-delivered beam remains unchanged as a function of ion acceleration.  At this coherence level, the integrated beam path renders the ion-laser interaction immune to even very strong vibrations and the corresponding time-varying Doppler shifts; the peak Doppler shift corresponding to the highest acceleration measured here is ${\sim}9$~kHz.  This suggests that coherence limited by such perturbation, for instance in a fieldable sensor or clock platform, may be improved by using integrated photonics for quantum control of trapped-atom systems.

Recent demonstrations of the integration of ion-control technologies show the promise of ion arrays for scalable quantum information processing~\cite{mehta2016integrated,slichter2017uv,PhysRevApplied.11.024010, mehta2020int2Qgate}.  We have implemented full photonic integration of all the visible and infrared wavelengths required to ionize, cool, state-prepare, coherently control, and detect Sr$^{+}$ ions---an important milestone toward the development of practical quantum information processors with trapped ions. Furthermore, the delivery of light directly from fibers to an ion-trap chip was shown to provide the additional benefit of vibration-resilient coherent quantum control, which may enable a new class of robust and portable ion-trap-based clocks and quantum sensors deployable in environments beyond the laboratory.

We note complementary work~\cite{mehta2020int2Qgate} demonstrating efficient on and off-chip coupling of red light and its use to demonstrate high-fidelity multi-qubit operations.  Together with our work, these demonstrations provide further compelling evidence for the promise of integrated photonic technology for quantum information processing in atomic systems.

\bibliographystyle{naturemag}
\bibliography{ref}


\section{Methods}

\subsection{Trap-chip fabrication}

Fabrication of the trap chips begins with an $8$-inch silicon wafer on which 1~$\mu$m of SiO$_{2}$ is deposited via plasma-enhanced chemical vapor deposition (PECVD).  A $0.5$-$\mu$m-thick sputtered Nb metal layer is then deposited and patterned via optical lithography to form a ground plane for the ion trap~\cite{mehta2014CMOStrap}.  Next, a $5$-$\mu$m-thick layer of SiO$_2$ is deposited, again using PECVD, to form the lower cladding of the waveguides.  A $100$-nm-thick layer of SiN is then deposited via PECVD, patterned via optical lithography, and fully etched to form the waveguide cores.  The gratings are patterned in a subsequent optical lithography step and are then formed via a partial ($40$-nm-deep) etch of the SiN.  Another $5$-$\mu$m-thick layer of SiO$_2$ is then deposited via PECVD above the SiN to form the top cladding of the waveguides.
The SiN and SiO$_2$ have indices of refraction of $n_\text{SiN}=1.89$ and  $n_{\text{SiO}_2}=1.5$ measured at $\lambda=633$~nm.
Subsequently, a $1$-$\mu$m-thick Nb layer is deposited, patterned via optical lithography, and etched to define the trap electrodes and open the square apertures above the grating couplers. Finally, a $20$-nm-thick layer of ITO is deposited and patterned over the apertures in the trap metal to mitigate the potential charging of exposed dielectrics directly below that might otherwise compromise trap stability~\cite{wang2011laser}.   In this trap, we measured a single-ion axial mode heating rate of 640~$\pm$~200~quanta/s
with the ion located at the center of the trap and an axial trapping frequency of 1.4 MHz.  The measured heating rate is high compared to traps of similar design, but made from a single metal layer on a sapphire substrate, operated cryogenically; in such traps we have measured approximately 10~quanta/s under similar conditions~\cite{PhysRevA.91.041402}.  We have not yet determined if this is due to the presence of the ITO or to some other specific aspect of the processing used to fabricate these devices, such as the fiber-array attachment or the thick SiO$_2$ layer separating the electrode-metal layer from the waveguide layers; this SiO$_2$ layer is present in the gaps between the electrodes and hence only partially shielded by the electrode metal.

\subsection{Grating coupler design and performance considerations}
\label{gratingmeth}
The gratings are designed with a uniform period so that they do not focus in the vertical transverse direction (the direction perpendicular to the beam and closest to $z$) and maintain ${\sim}11$~$\mu$m beam diameters ($1/e^{2}$ intensity) at the ion location. 
However, the gratings are designed with a slight curvature to focus the beams in the horizontal transverse direction (the direction perpendicular to the beam and parallel to the chip surface) to ${\sim}4$--$7$~$\mu$m beam diameters.

\begin{figure*}[t]
\centering
\includegraphics[width=0.99\textwidth]{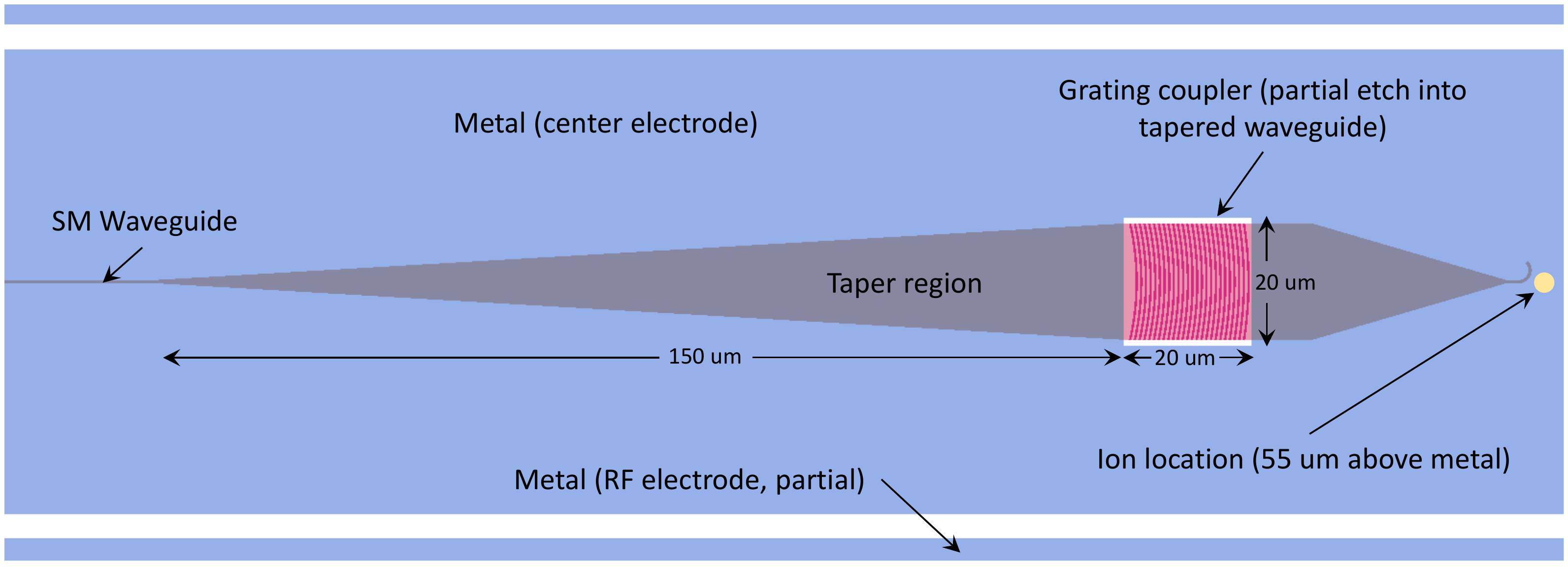}
\caption{Plan view of one of the grating couplers superimposed on the trap electrode metal (partial).  The single-mode (SM) waveguide is tapered to provide a wider beam at the grating coupler.  The grating teeth are formed via a partial etch into the waveguide material beneath a window in the electrode metal.  The gratings used in this work are forward-emitting, designed to emit a beam to the ion location as depicted, but $55$~$\mu$m above the surface of the metal.
}
\label{fig:grating}
\end{figure*}

Light launched into the single-mode waveguides was in all cases linearly polarized in the plane of the chip, and the quasi-transverse-electric-field mode excited in the waveguides maintains this polarization in the plane of the chip and perpendicular to the propagation direction.  The polarization is maintained  due to the substantial birefringence in the waveguides, which arises from the difference of the spatial modes associated with orthogonal polarizations of the light. Due to the symmetry of the waveguide, waveguide taper-region, and grating, and their relative configuration (see Figure~\ref{fig:grating}), the grating couplers emit light with polarization predominantly parallel to the chip, as well.  While the polarization purity was not measured here, we expect it to be high, as gratings of very similar design have been shown to emit light with a linear-polarization impurity of ${<}3\times10^{-4}$ in relative intensity~\cite{mehta2017precise}.  The applied magnetic field is nominally perpendicular to the chip surface at the ion location, and hence is also perpendicular to the light polarization, but its precise angle was not measured.

The grating couplers used in this work were designed using indices of refraction associated with low pressure chemical vapor deposition (LPCVD) SiN, a material known to have low optical loss, especially for infrared wavelengths~\cite{Bauters:11}.  However, as described above, PECVD SiN was chosen for fabrication of the devices.  This choice was made for two reasons.  First, in the time between grating design and the fabrication of the chips in this work, we developed PECVD SiN with substantially lower propagation loss at wavelengths below 460~nm, as compared with LPCVD SiN.  Second, since only the PECVD SiN can be deposited at temperatures compatible with pre-existing metal layers on the wafers, using PECVD SiN enables us to place a continuous metal ground plane between the integrated photonics and the Si substrate; we believe that doing so leads to more stable operation of the ion trap due to the shielding of the Si from stray light from the photonics, which could otherwise generate photoelectrons in the Si and lead to fluctuations of the chip's electrical impedance and thus the RF voltage delivered to the trap.

The trade-off in choosing PECVD SiN is that, due to the indices of refraction differing from those of LPCVD SiN, the grating emission angles deviate significantly from the original targeted values. Had we assumed the indices of the PECVD SiN in our grating design, the light emitted from the gratings would have been much closer to our target location.  We expect that residual deviations from target when assuming the proper indices of refraction will be dominated by our uncertainty in these indices, as our simulations have shown us that the misalignment is most sensitive to the tolerance on the index.  As a result, the optical index of the grating material is the primary property that must be controlled in fabrication in order to ultimately achieve small alignment errors. Other fabrication tolerances, such as material thicknesses and the discrepancy between the size of features drawn on a lithography mask and what is fabricated, are important but less of a concern given a lower misalignment sensitivity to them, as well as what we believe to be our better ability to control these tolerances.

Another design choice was to make the gratings emit light in the forward direction and at angles far from normal to the chip surface. Forward emission in general makes the waveguide routing considerably simpler since backwards-emitting gratings located in the positions used in our trap design would necessitate crossing of the photonics pathways given the finite bending radii of the waveguides.  The far-from-normal emission angle also helps in a similar regard with respect to routing, as it allows the gratings to be moved further from the ion location, thus providing the necessary space for the grating couplers to be placed.  

The choice of forward emission at the angles desired here also comes with a trade-off.  Higher-order modes of diffraction can be generated, sending light to spurious locations above the chip which could interfere with quantum operations.  In order to greatly reduce higher-order diffraction, we chose to design the gratings with a partial etch of the grating teeth at the expense of a slightly reduced grating efficiency into the first-order diffracted beam of interest.  The reduction of grating-emission efficiency into all orders is a result of the reduced index contrast between the partially-etched SiN teeth and the SiO$_2$ gaps between the teeth, as compared with the contrast that would exist for fully-etched SiN.

\subsection{Integrated photonic beam characterization}

\begin{figure*}
\centering
\includegraphics[width=0.9\textwidth]{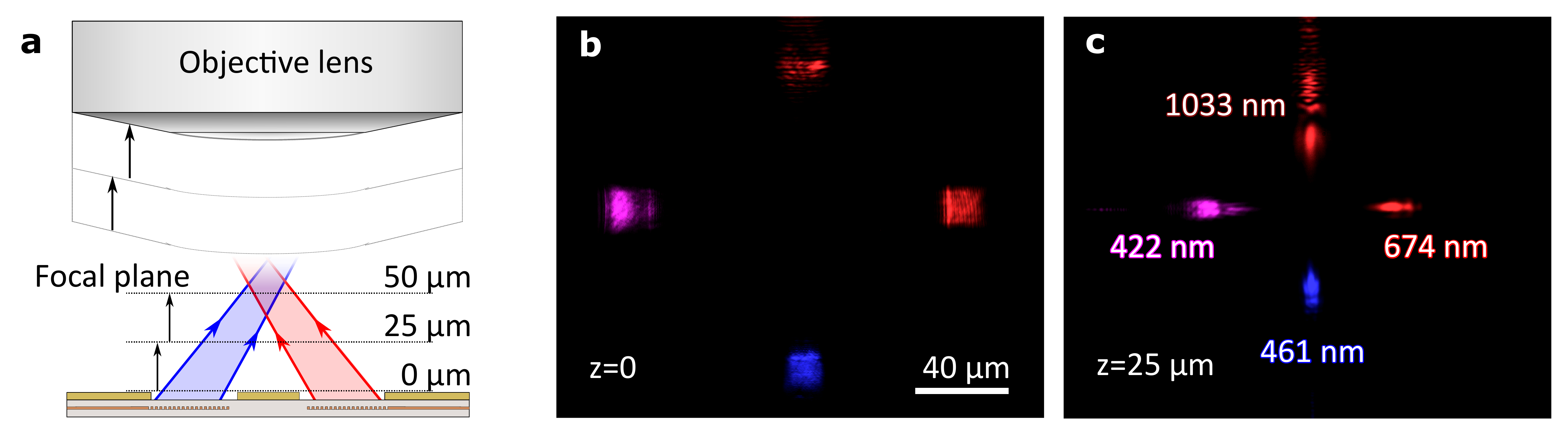}
\caption{Integrated photonic beam profiles measured from camera-recorded images.
(a) High-NA microscope images of the beams are taken while vertically scanning the focal plane above the chip. 
Laser light is emitted from the grating couplers and imaged at a height of $z=0$ (b) and $z=25$~$\mu$m (c) above the ion trap electrodes.
}
\label{fig:imageprofileSup}
\end{figure*}

To profile the integrated beams outside of the ion trap vacuum system, we use a custom-built, high-NA beam profiler.  This profiler consists of an infinity-corrected 0.9 NA, 60X (magnification) microscope objective and a 1X tube lens, followed by a CMOS-sensor-based camera located at the image plane of the lens system.  The lenses serve to translate the profile of the beam from the objective plane to the detection sensor (magnified by the lens system's magnification).  The high-NA system allows us to profile beams that are both rapidly diverging, due to tight foci, and travelling at large angles relative to the system's optical axis.  To profile the beams as a function of height above the trap chip, we step the height of the profiler above the surface using a stepper-driven translation stage with ${\sim}1$-$\mu$m precision and measure the beam profile as a function of the vertical position (Figure \ref{fig:imageprofileSup}). Combining these beam cross sections allows the reconstruction of the full three-dimensional beam profile, providing the beam diameter, focus height, and angle of emission.

\subsection{Beam profiling via \textit{in situ} ion measurements}

We profiled the 674-nm integrated beam by measuring the frequency of Rabi oscillations on the qubit transition as a function of the ion's position along the trap axis~(see~\figRef{fig:profile}b). The ion-interaction profile resolves some of the non-Gaussian beam structure evident in the line cut of the laser intensity from images taken with the microscope \textit{ex situ} and shows a beam diameter ($1/e^2$ intensity) of 13 $\pm$ 0.7 $\mu$m  along the axial direction, which is offset to the right of the trap center ($y=0$) at $y=13~\mu$m (see \figRef{fig:profile}c). 

The 422-nm integrated beam was profiled by measuring the fluorescence of an ion illuminated by 422-nm light emitted from the grating and a free-space 1092-nm beam.  In these measurements, the beam intensities were kept below saturation so that the ion fluorescence rate was approximately proportional to both the 422-nm and 1092-nm intensity.  We measure a beam diameter of 8.5 $\pm$ 1 $\mu$m  along the axial direction, offset to the left of the trap center at $y=-11~\mu$m (\figRef{fig:profile}d). While both the 422-nm and 674-nm beams are focused transversely along $x$, profiling with the ion along $y$ measures the unfocused cross-section.  

The 1092-nm integrated beam was profiled by measuring ion fluorescence while illuminated by a free-space 422-nm beam (again using intensities below saturation) and shows that the beam focuses along the axial direction ($y$) to a beam diameter of 5.5 $\pm$ 0.25 $\mu$m (\figRef{fig:profile}e).  We profiled this same infrared integrated beam path using 1033-nm light.  First we prepared the ion in the upper (dark) state by applying a $\pi$ pulse with an external 674-nm beam. Then we applied a short quench pulse of 1033-nm light via the integrated beam path to partially quench the dark state population. 
This increased the probability for the ion to be in the lower qubit (bright) state but maintained finite dark state probability (to avoid saturation). Finally, we measured the probability that the ion was in the bright state at various positions along $y$ and obtained a beam diameter of 6.7 $\pm$ 0.3 $\mu$m (see \figRef{fig:profile}f).

\begin{figure}[t]
\centering
\includegraphics[width=0.4\textwidth]{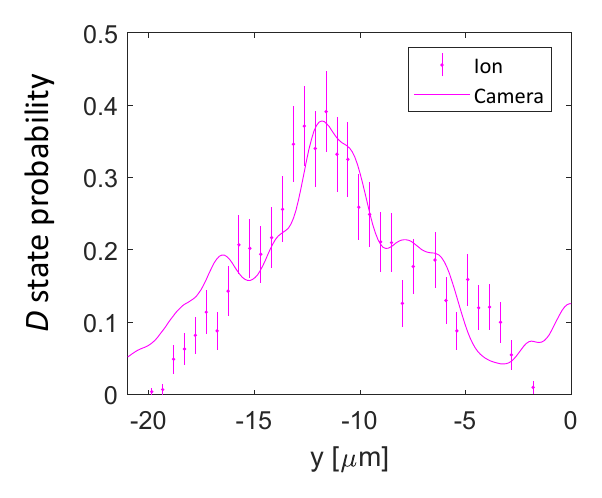}
\caption{Ion interaction profile of 408-nm light, which was used as a proxy for 405~nm.  Error bars indicate the standard error of the mean.}
\label{fig:ion408}
\end{figure}

Whereas Sr photoionization was verified using 405-nm light emitted from a grating coupler designed for blue light coupling (nominally 422~nm), ion loading using this method did not provide sufficient statistics to allow precise beam-mode characterization. Instead, light near 408~nm, which drives the $S_{1/2} \rightarrow P_{3/2}$ transition, was used as a proxy for 405~nm.
An ion excited to $P_{3/2}$ has a small probability of decaying to the metastable dark $D_{5/2}$ state. Therefore the probability the ion is in the upper qubit (dark) state is proportional to the intensity of the 408~nm at the ion location. 
To avoid saturation, we chose an input laser power and detuning so that when the ion is in the center of the 408 nm beam, it is in the dark state ${\sim}40\%$ of the time. The integrated violet beam angle is such that the beam center is displaced to $y=11.4$~$\mu$m and has a beam diameter of 11.3 $\pm$ 1.3 ~$\mu$m along the axial direction (see Figure~\ref{fig:ion408}).

All the ion interaction profiles agree very well with the beam profiles measured using the microscope objective, which gives us confidence that the integrated photonic components can be accurately characterized, independent of the ion, using conventional optical techniques and equipment. In addition, this implies that there are not significant effects on the integrated optics from cryogenic or UHV operation.

\subsection{Ion State Measurement}

When an ion in the $D$-state decays to the $S$-state during the detection interval, it will start emitting photons  (i.e. a dark ion becomes bright), leading to measured photon counts which follow neither the dark nor bright histograms.  Hence, the total decay probability during the detection interval overestimates the error in misidentifying a dark ion as a bright one.  To accurately determine the mean qubit detection fidelity, we extract $\epsilon_{d}$ and $\epsilon_{b}$ by examining the overlap of the bright-state histogram with the expected dark-state histogram with decay, calculated using the measured bright and background  histograms, as follows.  We integrate, for all decay times $t$, the total probability to register any particular number $k$ of photons (a sum over the probability to measure $j$ photons from the background histogram and $k-j$ photons from the bright histogram, for $j\leq k$) in the detection interval, weighted by the probability to decay at time $t$.  The latter is given by the exponential distribution, with decay time 390~ms for the $D_{5/2}$ state in ${}^{88}$Sr$^{+}$.

For the demonstration of ion-state readout using 422-nm and 1092-nm beams, both emitted from integrated optics, described in the main text, detection was affected by background photon scatter.  With spatial filtering of background photons, predominantly scattered from the 422-nm diffraction grating itself, we achieve count rates of $4540$~s$^{-1}$ from the ion (reduced from typical collection rates due to the additional spatial filtering) compared to $967$~s$^{-1}$ due to background (which is dominated by unfiltered scattering from the grating).

Besides the above state readout measurement, we also demonstrated a second, similar measurement using a free-space 1092-nm beam in combination with the integrated 422-nm beam path.  The fidelity was measured in the same manner as for the detection method described in the main text, once again considering decay during measurement as described above.  In this case, we measure an expected mean optical-qubit detection fidelity of 99.6\% in 3~ms, an improvement due mainly to a higher scattering rate from the ion when in the $S$-state when compared to the integrated-optics-only measurement, which suffered from lower combined beam intensity at the ion location. A variation on this mode of detection, in which the 422-nm light is delivered locally to ions in an array while the 1092-nm light is delivered globally via free-space, may be preferred in some applications; the required 1092-nm repumping intensity is quite low, and in general the constant presence of light at this wavelength at all array sites is minimally detrimental as it does not couple strongly to either of the optical-qubit states.

\subsection{Qubit Coherence Measurements via Ramsey Interference}
In order to measure the qubit coherence time, first a $\pi/2$~pulse is applied using 674-nm light to create an equal superposition in the qubit; this is followed by a variable delay time; next, another $\pi/2$ pulse of a varying phase relative to the initial pulse is applied; finally, the probability that the ion is in the lower qubit (bright) state is measured via resonance fluorescence.  The bright state probability versus $\pi/2$ pulse phase is then fitted to a sinusoid and the Ramsey fringe contrast is extracted.  The Ramsey contrast versus delay time is subsequently fitted to a decaying exponential and the 1/$e$ qubit coherence time is determined.

To characterize the vibrational resilience of light delivered via integrated optics, we measured ion-qubit Ramsey interference as a function of vibrational coupling to the cryocooler.  To determine the ion acceleration with respect to the optical table when intentionally introducing vibrations, we imaged the ion fluorescence using an optical system mounted to the table.  Since the ion was in general presumably oscillating in three dimensions, we bounded the apparent ion acceleration's projection in two dimensions by monitoring the ion motion on an electron-multiplying CCD camera at high frame rate (${\sim}1000$~frames per second).  The ion excursion in space as a function of time was measured to extract a velocity, and using an approximation of circular motion to provide a lower bound on the peak acceleration, we calculated the acceleration and Doppler shift from the velocity and displacement.  The dominant oscillatory motion of the ion had a period of 15~ms, corresponding to a vibrational frequency of the coupled cryocooler-vacuum-chamber structure at 67~Hz (the Gifford-McMahon cycle of the cryocooler is driven at 1.2~Hz, but we conclude that the structure is resonant at this higher observed frequency).

While the decay with acceleration when using the integrated interrogation beam is consistent with 0 (see Fig.~\ref{fig:vibration}), we can bound the achieved vibration-isolation factor by comparing the uncertainty in the fitted decay constant for these data to the value of the fitted decay constant for the data taken with free-space interrogation.  This suggests that an  acceleration suppression factor of at least 25 is afforded by chip-integration of light-delivery optics.

\subsection{On-Chip Coupling and Total Optical Loss}

To maximize optical input coupling efficiency, the edge of the chip where coupling occurs is optically polished after dicing. 
In addition to the waveguides that are designed to deliver light to the ion, we use a ``loop-back" waveguide to monitor coupling efficiency into the chip as we align the fiber array to the waveguide inputs.  The fiber array containing six polarization-maintaining single mode optical fibers is aligned by optimizing the optical power through the loopback path, which we found to be a straightforward method to approximately optimize the optical power through all four input waveguides that deliver light to the ion. The block is then attached to the chip in a two-step process, first with UV-curable epoxy and then with cryogenic-compatible epoxy to provide additional structural support.

All beams are coupled into the waveguides from fiber with polarization oriented parallel to the surface of the chip. Before mounting the chip into the vacuum chamber, the power output from each grating was measured relative to the power of the light in each fiber to determine the total fiber-to-output (FTO) loss.  The SiN waveguide propagation losses were determined independently using separate test chips, co-fabricated with the ion trap chips, in which the transmission of light through waveguides of varying length was measured to extract the loss per waveguide length.  On-chip coupling (OCC) loss was also determined using separate co-fabricated devices that each consisted of a continuous waveguide running between opposing facets of the test chip and having input and output tapers identical to the input tapers used in the ion trap chip. Fibers (identical to those used for coupling into the ion trap chip) were placed at the input and output facets of the test device and the light transmission was measured.  By subtracting off the propagation loss and dividing this result by a factor of two, we extract the OCC loss.  The grating loss was then determined by subtracting the propagation loss and the OCC loss from the FTO loss.  

The SiN waveguides have propagation losses below 0.5~dB/cm for wavelengths above 633~nm, with losses increasing to ${\sim}10$~dB/cm at 405~nm (for more details see ref.~\cite{sorace2019versatile}).  For the 674-nm integrated optical path, we obtain an FTO loss of $21.4$~dB, which includes the OCC loss ($10$~dB), propagation loss in the routing waveguide ($0.4$~dB for approximately 0.75~cm), and loss of the grating coupler ($11$~dB).  
For the 422-nm path, we measure $10$~dB OCC loss, $3$~dB propagation loss, and $12$~dB grating coupler loss.
For 461~nm we measure $11$~dB OCC loss, $1.5$~dB propagation loss, and $9$~dB grating coupler loss.
For 1092~nm we measure $6$~dB OCC loss, $0.4$~dB propagation loss, and $10$~dB grating coupler loss.

When the ion-trap chip is mounted in the vacuum system, we observe an additional $3$~dB of loss arising from fiber splices and fiber-to-fiber connectors that are part of the fiber vacuum feedthrough.  During cool-down from room temperature to 7~K, the optical alignment of the fiber block deteriorates, resulting in further loss of approximately $7$~dB in the 422-nm, 461-nm, and 674-nm beam paths. These additional losses are determined by directly measuring the light emitted from the gratings through windows in the vacuum system. (The additional loss in the 1092-nm beam could not be accurately measured due to difficulty in measuring this light outside of the vacuum system.)  This increases the total loss of the 674-nm path to $31.4$~dB, the 422-nm path to $35$~dB, the 461-nm path to $31.5$~dB, and the 1092-nm path to $26.4$~dB, where we have assumed that the 1092-nm loss increases by the same $10$~dB as observed for the other beams. In the case of the 674-nm light, our measurement of a 6.5 $\mu$s $\pi$ pulse time agrees with the value we obtain through a first-principles calculation of the Rabi frequency~\cite{roos2000controlling} given the measured beam dimensions, the beam power input to the chip, and the total optical loss.

\begin{table*}[tb]
\begin{center}


    \begin{tabular}{|c|c|c|c|c|c|c|}
    \hline
        Wavelength & On-Chip Coupling & Propagation & Grating & Fiber Feedthrough & Cooldown & Total  \\ \hline \hline
        422 nm & 10 dB & 3 dB & 12 dB & 3 dB & 7 dB & 35 dB \\ \hline
        461 nm & 11 dB & 1.5 dB & 9 dB & 3 dB & 7 dB & 31.5 dB \\ \hline
        674 nm & 10 dB & 0.4 dB & 11 dB & 3 dB & 7 dB & 31.4 dB \\ \hline
        1092 nm & 6 dB & 0.4 dB & 10 dB & 3 dB* & 7 dB* & 26.4 dB* \\  \hline
    \end{tabular}
    
    
\end{center}
\caption{Summary of coupling and on-chip optical losses versus wavelength.  *The 1092-nm entries for fiber-feedthrough, cooldown, and total loss were not measured directly, but rather were inferred from measurements of the other beams.}
\label{losstable}
\end{table*}

A summary of the optical losses is given in Table \ref{losstable}.  The dominant sources of optical loss here are the on-chip coupling and grating-coupler loss.  The on-chip coupling loss arises due to the optical-mode-size mismatch between that of the fibers and the integrated waveguides.  This could be improved by expanding the mode in the waveguides, e.g.,  via thinning or further lateral tapering of the SiN core or via the use of an input coupler that has a waveguide core material with lower optical index.  In these cases, the effect is to provide weaker confinement of the light in the waveguides so that the mode size is increased.

The grating coupler loss arises primarily from the finite index contrast between the SiN grating teeth and the SiO$_2$ gaps between these teeth. As noted earlier,  the index contrast is further reduced by our choice to only partially etch the SiN. The grating coupler efficiency could in principle be increased by simply increasing the length of the grating section, thus providing a larger region for diffraction.  However, as space to place the couplers around the ion location is already at a premium, increasing the size of the gratings is not very practical for this application.  A method to increase the grating-coupler efficiency for a fixed lateral extent of the grating is to use multiple grating layers to generate constructive interference for light emitted upwards from the grating.  This method requires more complex design and fabrication and was not pursued in this work.  Another technique is to generate constructive interference using the downward-diffracted light that is subsequently reflected upwards from the silicon substrate~\cite{mehta2017precise, mehta2020int2Qgate}. However, this is a less flexible approach, especially when using multi-wavelength light, as is the case in this work.  This is because the requirement for constructive interference, in the presence of a fixed-thickness layer between grating and substrate, constrains the angle of light emission, and this constraint is wavelength dependent.

Despite these losses, we achieved light-ion interaction strengths that are comparable to those obtained when using free space beam paths with similar optical power because the cross-sectional area of the beams in the free-space case is typically over 100 times larger than that of the integrated beams. We believe that all of these loss channels can be significantly improved with new waveguide materials~\cite{west2019lowloss}, alternative packaging techniques~\cite{mehta2020int2Qgate}, and improved grating designs~\cite{mehta2017precise, mehta2020int2Qgate}.



\section*{Acknowledgements}


We thank Peter Murphy, Chris Thoummaraj, and Karen Magoon for assistance with chip packaging, and Patrick Hassett and Karen Yu for chip-facet polishing.  This material is based upon work supported by the Department of Defense under Air Force Contract No. FA8702-15-D-0001.  Any opinions, findings, conclusions or recommendations expressed in this material are those of the authors and do not necessarily reflect the views of the Department of Defense.

\section*{Competing Interests}

The authors declare that they have no competing financial interests.

\section*{Correspondence}
Correspondence and requests for materials
should be addressed to\\ R.J.N. (email: robert.niffenegger@ll.mit.edu), J.M.S. (email: jsage@ll.mit.edu), or\\ J.C. (email: john.chiaverini@ll.mit.edu).

\section*{Data Availability}
All relevant data are available from the corresponding author upon request.


\section*{Author Contributions}
J.M.S. and J.C. conceived of the work. C.S.-A. and S.B. designed the integrated optical components; D.K. oversaw the fabrication of the devices.  R.J.N. performed the experiments, with assistance from J.S., C.D.B., D.R., R.M., R.T.M., G.N.W., and W.L.; R.J.N. analyzed the data. All authors discussed the results and contributed to writing the paper.


\end{document}